\documentclass[preprint,12pt]{elsarticle}

\usepackage{graphicx}

\usepackage{amssymb}

\journal{Nuclear Physics A}

\begin{document}

\begin{frontmatter}



\title{Extended Longitudinal Scaling: direct evidence of saturation*}


\author{Wit Busza}

\address{Department of Physics and Laboratory for Nuclear Science, 
Massachusetts Institute of Technology, 77 Massachusetts Avenue, Cambridge, MA USA}

\begin{abstract}
Multiparticle production of charged particles at high energies exhibit the phenomenon of Limiting Fragmentation.  Furthermore, the region in rapidity over which the production of particles appears to be independent of energy, increases with energy.  It is argued that this phenomenon, known as Extended Longitudinal Scaling, is a direct manifestation of some kind of saturation, akin to that in the Color Glass Condensate picture of particle production.  


*Invited paper presented at the Glasma Workshop, BNL, May 2010
\end{abstract}


\end{frontmatter}

In 1969 Benecke, Chou, Yang and Yen \cite{1} proposed the ÒHypothesis of Limiting FragmentationÓ.  It was based on the ``two-fireball model'' used to explain cosmic ray data and the ``intuitive picture of a high-energy collision process as two extended objects going through each other, breaking into fragments in the process...''.  In essence, the Hypothesis of Limiting Fragmentation, which is a consequence of Feynman Scaling, states that if in the collision of two particles one views the collision in the rest frame of one of the incident particles, the production process of the soft particles is independent of the energy or rapidity of the other particle.  Manifestations of Limiting Fragmentation are apparent for example in figures 1-3, where pseudorapidity distributions of charged particles produced in p+p, p+A and A+A collisions at various energies are shown in the rest frame of one of the incident particles.  

\begin{figure}[htbp] 
   \centering
   \includegraphics[width=3.5in]{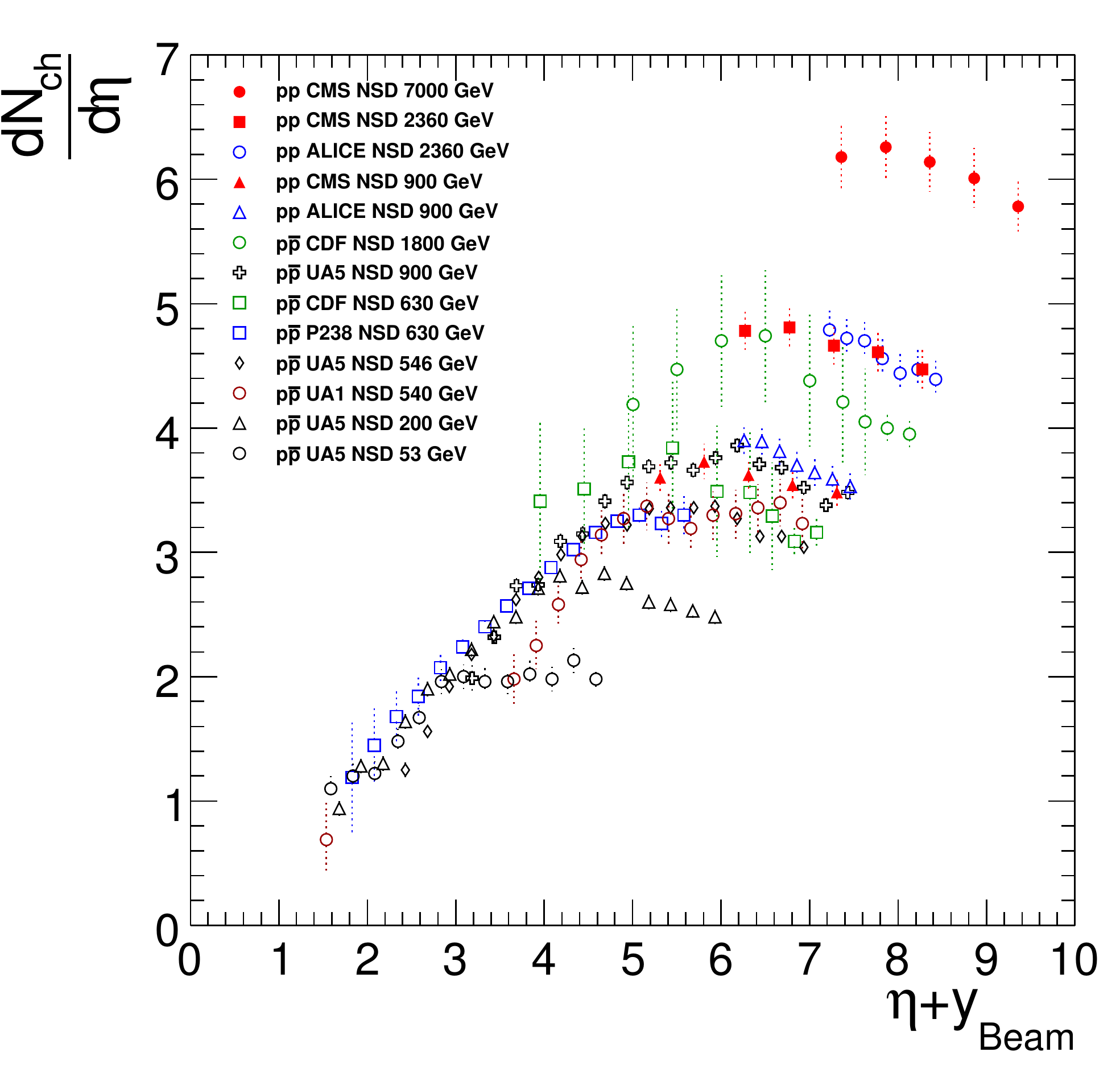} 
   \caption{(color on line) Example of Limiting Fragmentation and of Extended Longitudinal Scaling as seen in non-single diffractive (NSD) $p+p$ and $p+\bar{p}$ collisions.  $\frac{dN_{ch}}{d\eta}$ is the number of produced charged particles per unit of pseudorapidity $\eta$ plotted as a function of $\eta$ in the rest frame of the incident proton.  In boosting the data from the center of mass to the proton rest frame the difference in pseudorapidity and rapidity $y$ is ignored.  The figure is courtesy of Yen-Jie Lee \cite{2}.}
   \label{fig:1}
\end{figure}
\begin{figure}[htbp] 
   \centering
   \includegraphics[width=3.5in]{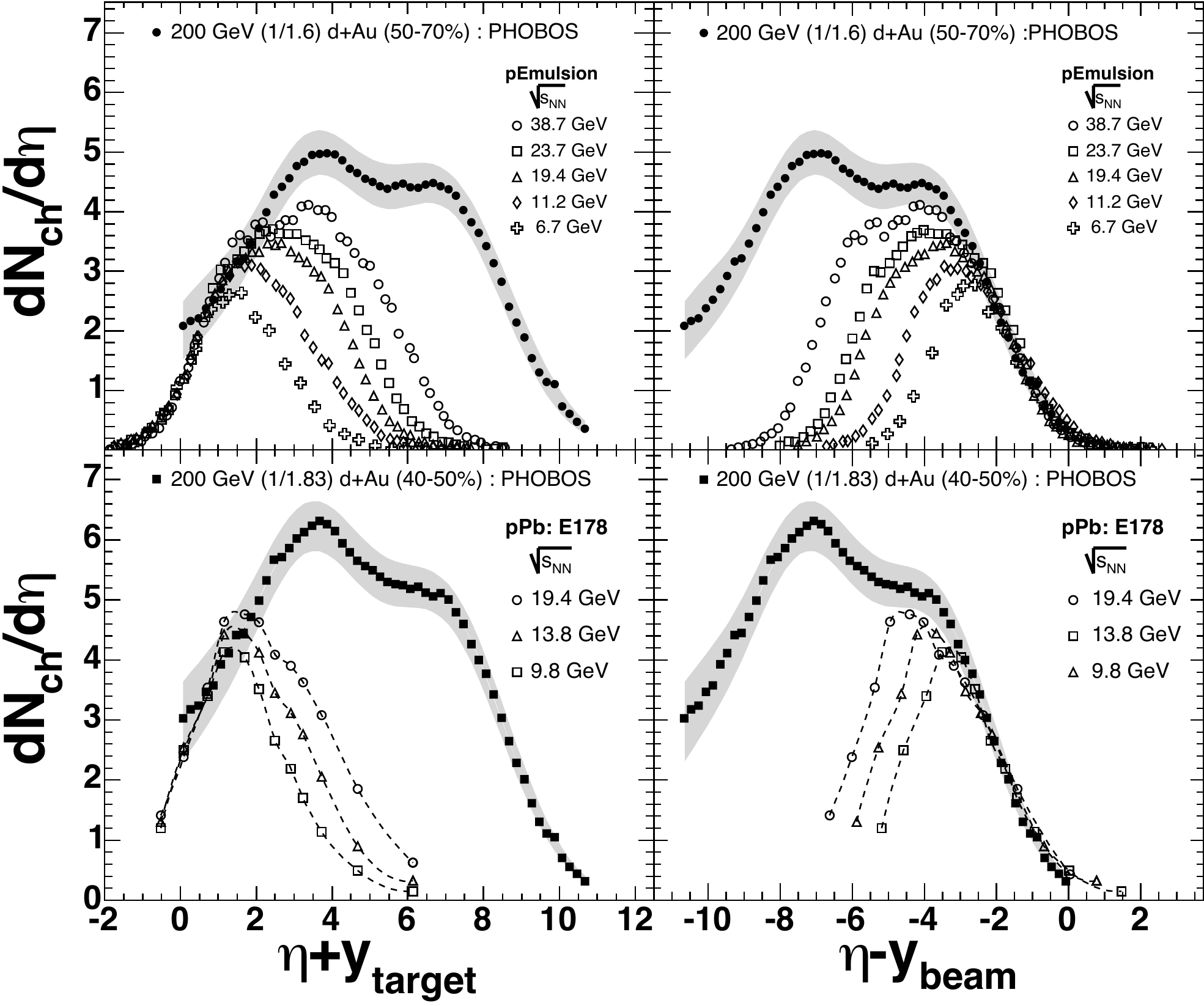} 
   \caption{Example of Limiting Fragmentation and of Extended Longitudinal Scaling seen in p+A and d+A collisions.  $\frac{dN_{ch}}{d\eta}$ is the total number of produced charged particles per unit of pseudorapidity $\eta$ plotted as a function of $\eta$ in the rest frame of one or the other incident particle.  In boosting the data from one frame to another the difference in pseudorapidity $\eta$ and rapidity $y$ is ignored.  The d+Au data are appropriately normalized to be comparable with the corresponding p+A data.  The figure is from \cite{3}.}
   \label{fig:2}
\end{figure}
\begin{figure}[htbp] 
   \centering
   \includegraphics[width=3.5in]{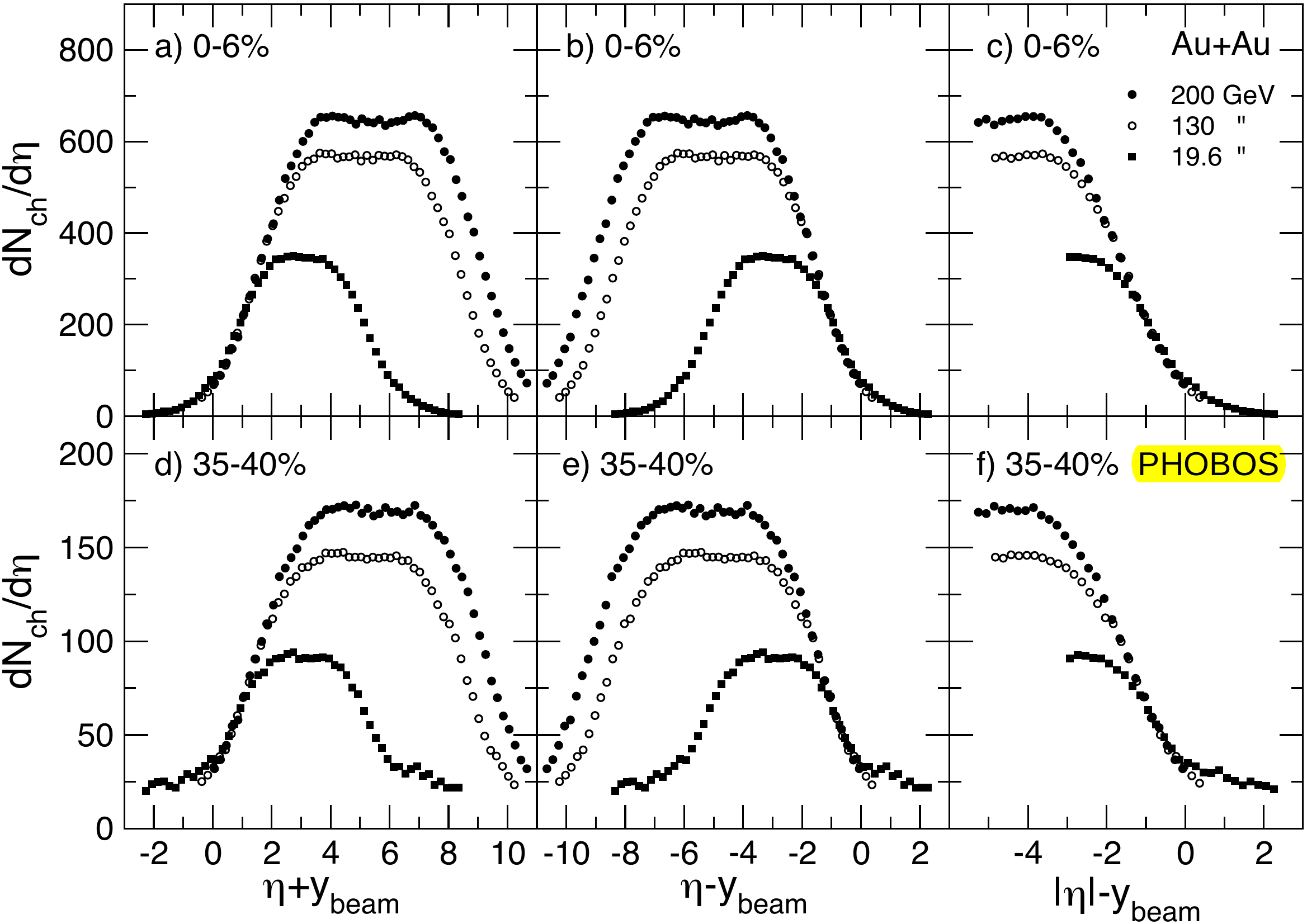} 
   \caption{Example of Limiting Fragmentation and of Extended Longitudinal Scaling seen in the most central (0-6\%) and more peripheral (35-40\% centrality) Au+Au collisions.  $\frac{dN_{ch}}{d\eta}$ is the total number of produced charged particles per unit of pseudorapidity $\eta$ plotted as a function of $\eta$ in the rest frame of one or other Au nucleus.  In boosting data from the center of mass frame to the Au rest frame, the difference in pseudorapidity $\eta$ and rapidity $y$ is ignored.  The figure is from \cite{3}.}
   \label{fig:3}
\end{figure}
It should be noted that throughout this paper no distinction is made between the pseudorapidity 
$\eta = -ln \tan \frac{\theta}{2} = \sinh^{-1}(\frac{p_{\|}}{p_{\perp}})$ and the rapidity $y= \sinh^{-1}(\frac{p_\|}{m_\perp})$, where $\theta$ is the polar angle of the produced particle, $p_\perp$ and $p_\|$ are respectively the perpendicular and parallel components of its momentum and $m_{\perp}^{2} =m^2 +p_{\perp}^{2}$ where $m$ is the mass of the produced particle.  

At high energies, to a precision relevant for the discussion here, the difference between $\eta$ and $y$ 
is not material.  
Examples of rapidity rather than pseudorapidity distributions which exhibit Limiting Fragmentation can be found in \cite{4, 5}. 

Figures 4-7 illustrate the universal nature of Limiting Fragmentation.  It is reflected not only in the energy independence of the rapidity distributions of particle densities but also of particle ratios, of centrality or A-dependence of particle production in A+A collisions, and even of $v_1$ and $v_2$, measures of the azimuthal unisotropy of particles produced in such collisions.
\begin{figure}[htbp] 
   \centering
   \includegraphics[width=3.5in]{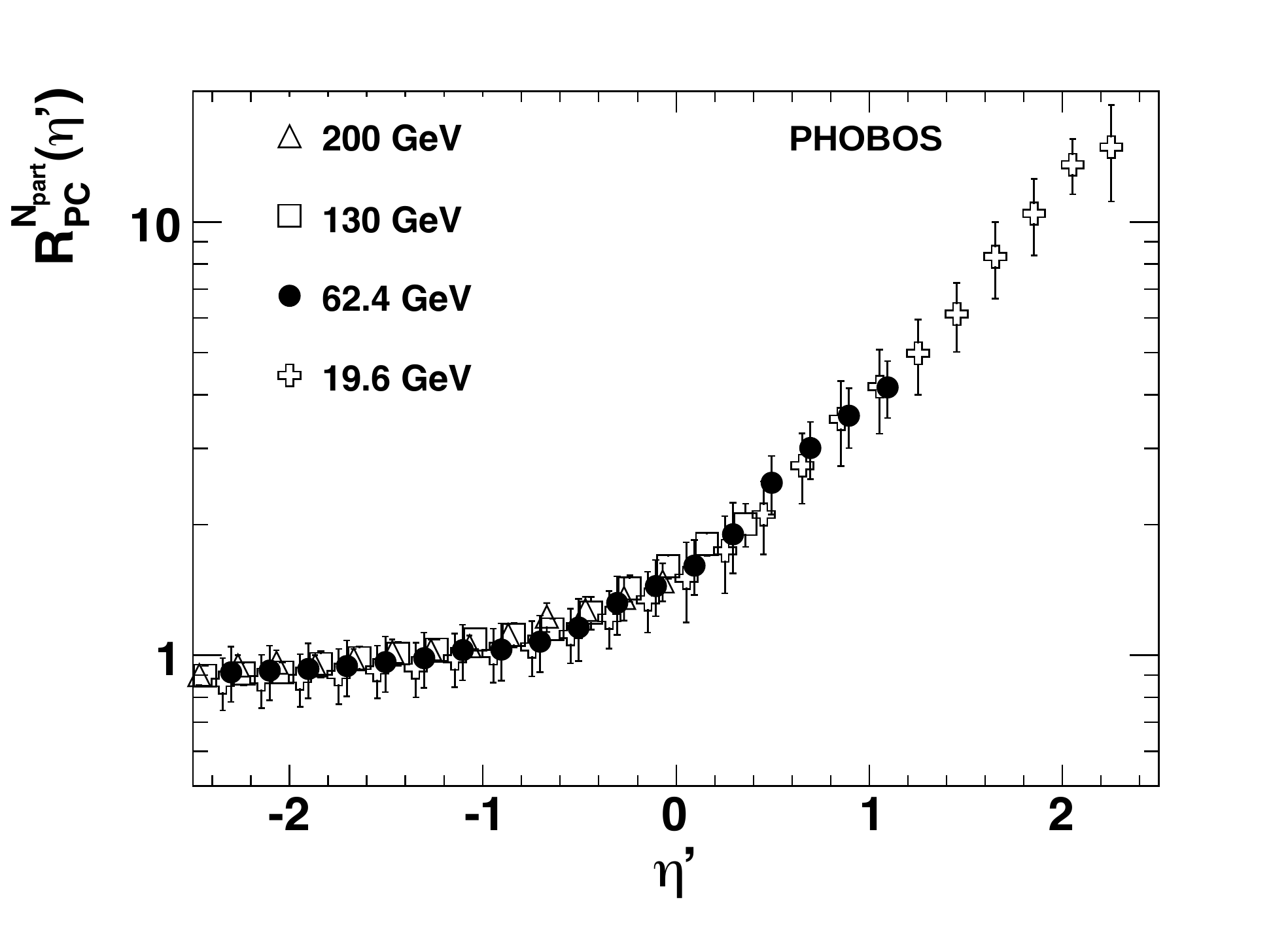} 
   \caption{Illustration of the degree to which Limiting Fragmentation seems to be exact in some processes.  In the figure the ratio of the number of charged particles produced in peripheral (35-40\% centrality) to central (0-6\% centrality) Au+Au collisions at various energies, normalized to the same number of participants $N_{part}$, is plotted as a function of the pseudorapidity $\eta^`$ of the produced particle, in the rest frame of one of the Au nuclei.  In boosting the data from the center of mass frame to the Au rest frame the difference in pseudorapidity and rapidity is ignored.  The figure is from \cite{6}.}
   \label{fig:4}
\end{figure}
\begin{figure}[htbp] 
   \centering
   \includegraphics[width=2.5in]{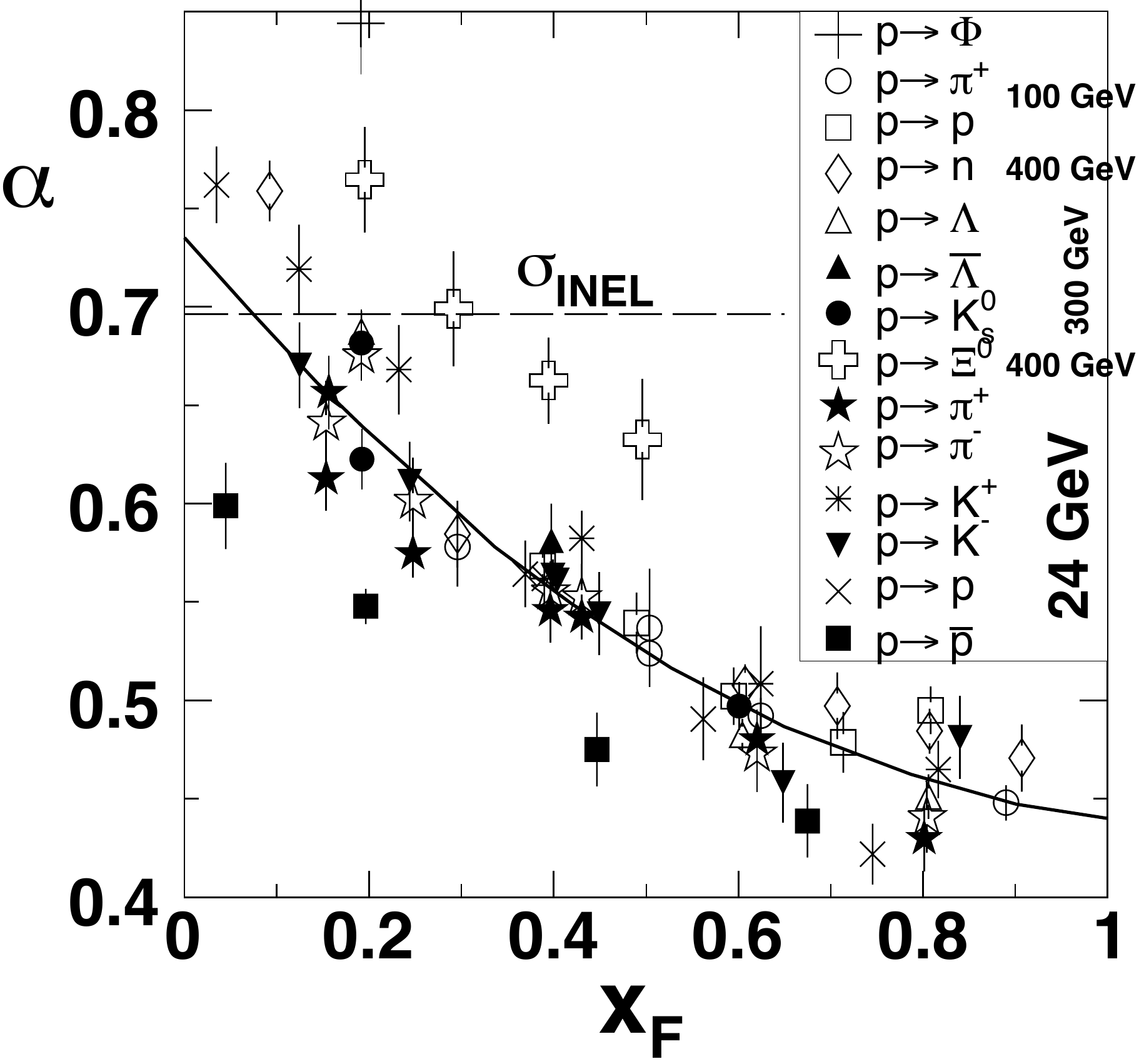} 
   \caption{Compilation of the A-dependence in the forward fragmentation region of the production of various particles in p+A collisions at a variety of energies.  $\alpha$ is the exponent in the parametrization of $p+A\to h+anything$ data in the form $\sigma_A=\sigma_oA^\alpha$.  It is plotted as a function of $x_F$, the ratio of the momentum of the hadron $h$ to the momentum of the incident proton.  All the data are for low $p_t (\le300\frac{MeV}{c}$).  The figure is from \cite{7}.}
   \label{fig:5}
\end{figure}
\begin{figure}[htbp] 
   \centering
   \includegraphics[width=3.5in]{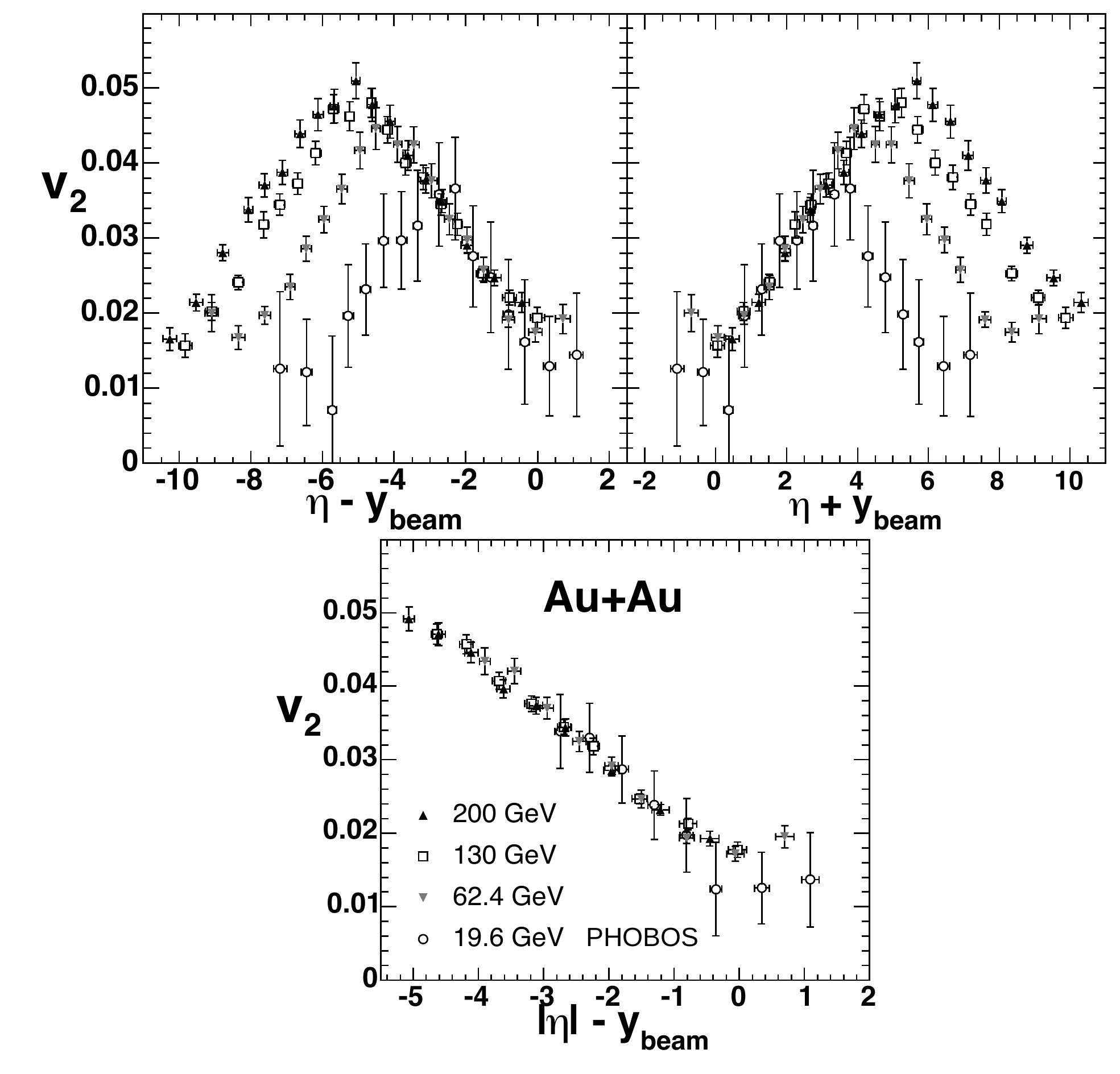} 
   \caption{Example of Limiting Fragmentation and Extended Longitudinal Scaling exhibited in the azimuthal distribution (in the so-called elliptic flow) of charged particles produced in Au+Au collisions.  $v_2$ is the second coefficient in the Fourier expansion of the azimuthal distribution of the charged particles.  In the top two panels $v_2$ is plotted as a function of the pseudorapidity in the rest frames of the two incident Au nuclei.  In boosting the data from the center of mass frame to the Au rest frame the difference in pseudorapidity and rapidity is ignored.  In the bottom panel the positive and negative rapidity data in the top panels are averaged.  The figure is from \cite{3}.}
   \label{fig:6}
   
\end{figure}
\begin{figure}[htbp] 
   \centering
   \includegraphics[width=2.5in]{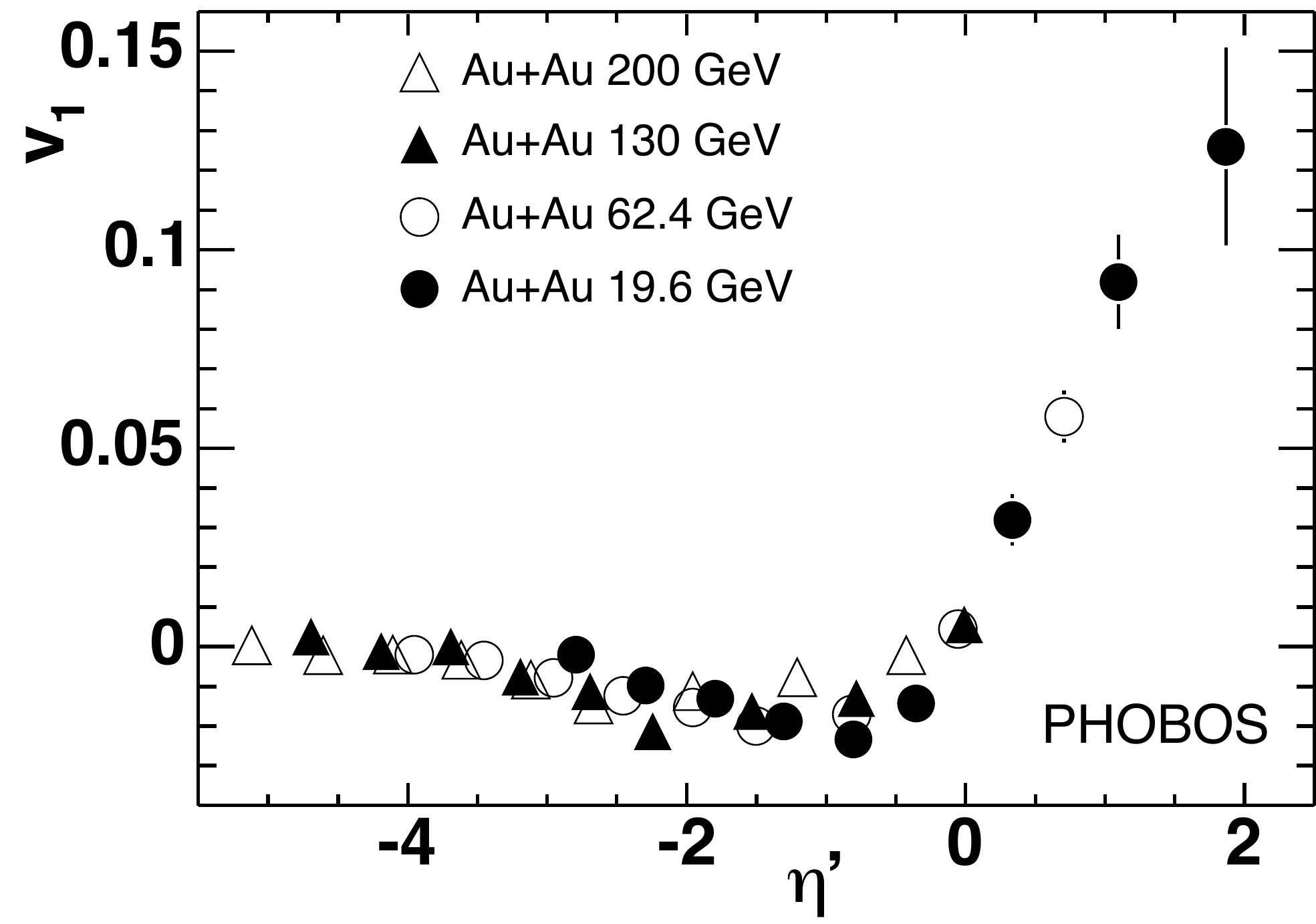} 
   \caption{Same as caption to figure 6 except that, in the data above, $v_1$ is the first Fourier coefficient.   The data show how directed flow exhibit Limiting Fragmentation and Extended Longitudinal Scaling.  The figure is from \cite{5}.}
   \label{fig:7}
\end{figure}

It should be pointed out that in fig 5 the independence of the A-dependence of the data on energy follows from Limiting Fragmentation.  However the values of the exponent $\alpha$ in the $A^\alpha$ parametrization of the particle density and the apparent independence of $\alpha$ on the nature of the produced particle is a new feature that, to the best of my knowledge is still not well understood.  Interestingly, it has characteristics which are strikingly similar to those of jet quenching.

Figures 1-7 show in the data another prominent feature related to Limiting Fragmentation.  It is that the region in rapidity, over which the production of particles appears to be independent of energy, increases with energy.  This ``increase in the region of Limiting Fragmentation'', first seen in p+A collisions in the 1970Õs \cite{8, 9, 10}, was rechristen a few years ago by Mark Baker of the PHOBOS collaboration as ``Extended Longitudinal Scaling''.   

The main goal of this brief paper is to point out that Extended Longitudinal Scaling is direct evidence that some kind of saturation takes place in high energy collisions.  

The easiest way for me to explain how Extended Longitudinal Scaling implies saturation is through a discussion of a ``gedanken experiment'', where the results of measurements are known from existing experimental data.  

Suppose we construct a Au+Au collider where the energies of the two beams in the laboratory frame of reference can be independently adjusted.  We call one of the two beams the ``target'' and the other the ``beam''.  The corresponding rapidity of the two beams in our laboratory frame we call $y_{``target"}$ and $y_{``beam"}$.  With this collider we proceed to study the dependence of particle production at $y=0$ (in our rest frame) as we change $y_{``target"}$ and $y_{``beam"}$.  

What will be the outcome of our measurements of $\frac{dN}{d\eta}\mid_{ y=0}$ the density of charged particles produced at $y=0$, and of $v_2\mid _{y=0}$, the measure of elliplic flow at $y=0$?  Examples of results that will be obtained in such measurements are shown in figures 8 and 9.

Figure 8 shows, for the most central (0-6\% centrality) Au+Au collisions, the $y_{``beam"}$ dependence of $\frac{\frac{dN}{d\eta}\mid_{ y=0}}{{(\frac{N_{part}}{2})}}$ for various values of $y_{``target''}$.  The data points shown are taken directly from figure 10, that gives the results of PHOBOS studies of Extended Longitudinal Scaling at RHIC.
\begin{figure}[htbp] 
   \centering
   \includegraphics[width=3in]{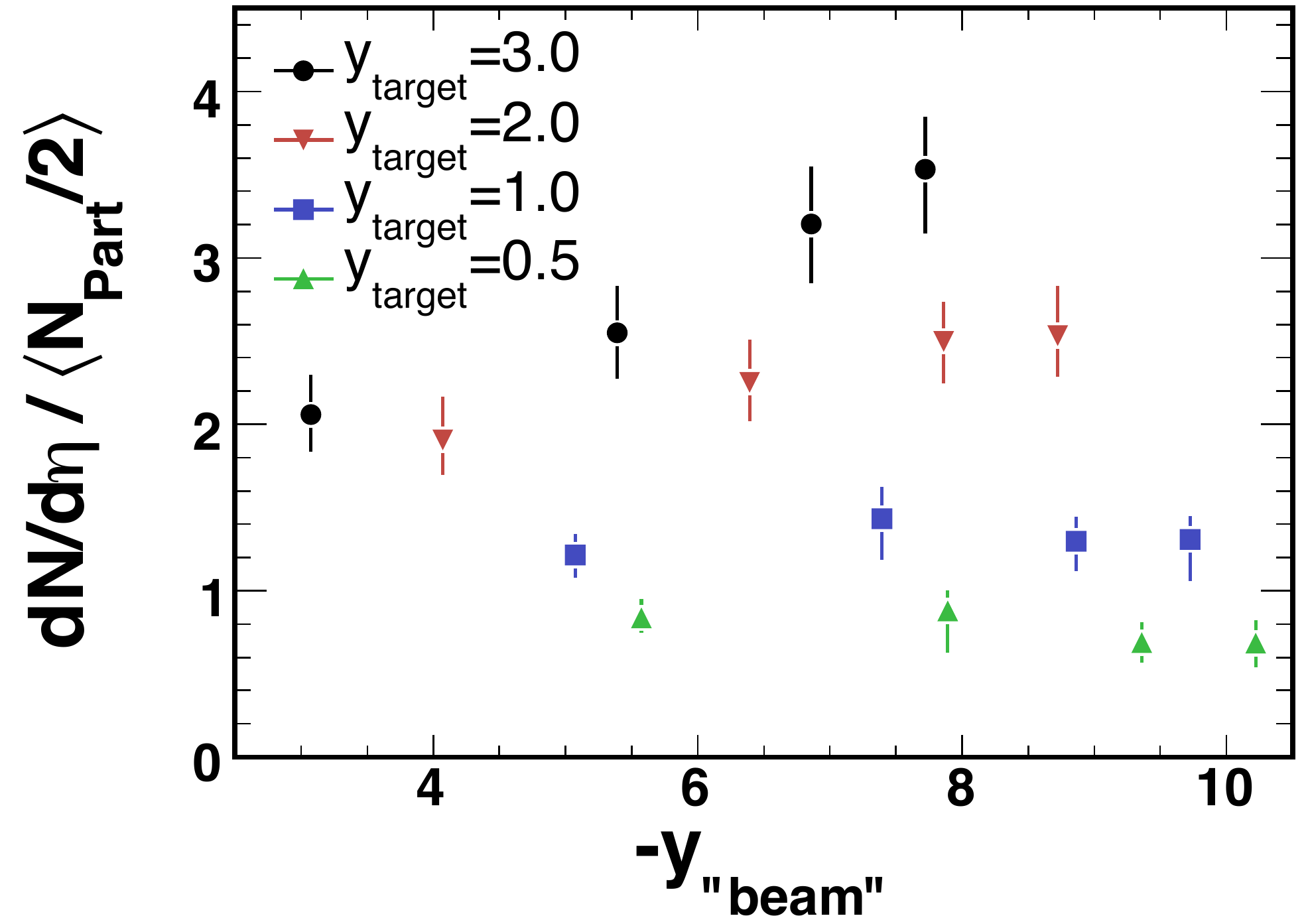} 
   \caption{(color on line) The $\frac{\frac{dN}{d\eta}\mid_{ y=0}}{{(\frac{N_{part}}{2})}}$ that would be measured in the ``gedanken experiment" (see text) if at $y=0$ in the laboratory frame of reference, a Au beam with a rapidity $y_{``beam"}$ collided with a Au beam with rapidity $y_{``target"}$.  The data show that, for a given value of $y_{``target"}$ the produced particle density reaches a saturated value as $y_{``beam"}$ increases.  Furthermore, that this saturated value is higher the larger is the the value of $y_{``target"}$.  The figure is from \cite{11}.}
   \label{fig:8}
\end{figure}

As an illustration, in our gedanken experiment results in fig 8, the value of$\frac{\frac{dN}{d\eta}\mid_{ y=0}}{{(\frac{N_{part}}{2})}}$ for $y_{``target"} = 2.0$ and $y_{``beam"} = 7.06$ is obtained from the PHOBOS measurement in figure 10 at $[|\eta|-y_{beam}]=-2.0$ for $\sqrt{s}=130GeV (y_{Au}=\pm 4.93)$.

Figure 9 shows, for the 0-40\% Au+Au collisions the $y_{``beam"}$ dependence of $v_2$ for various values of $y_{``target"}$.  In a similar manner in which the results shown in figure 8 are obtained from data in figure 10, the results in figure 9 are obtained from data in figure 6. 
\begin{figure}[htbp] 
   \centering
   \includegraphics[width=3in]{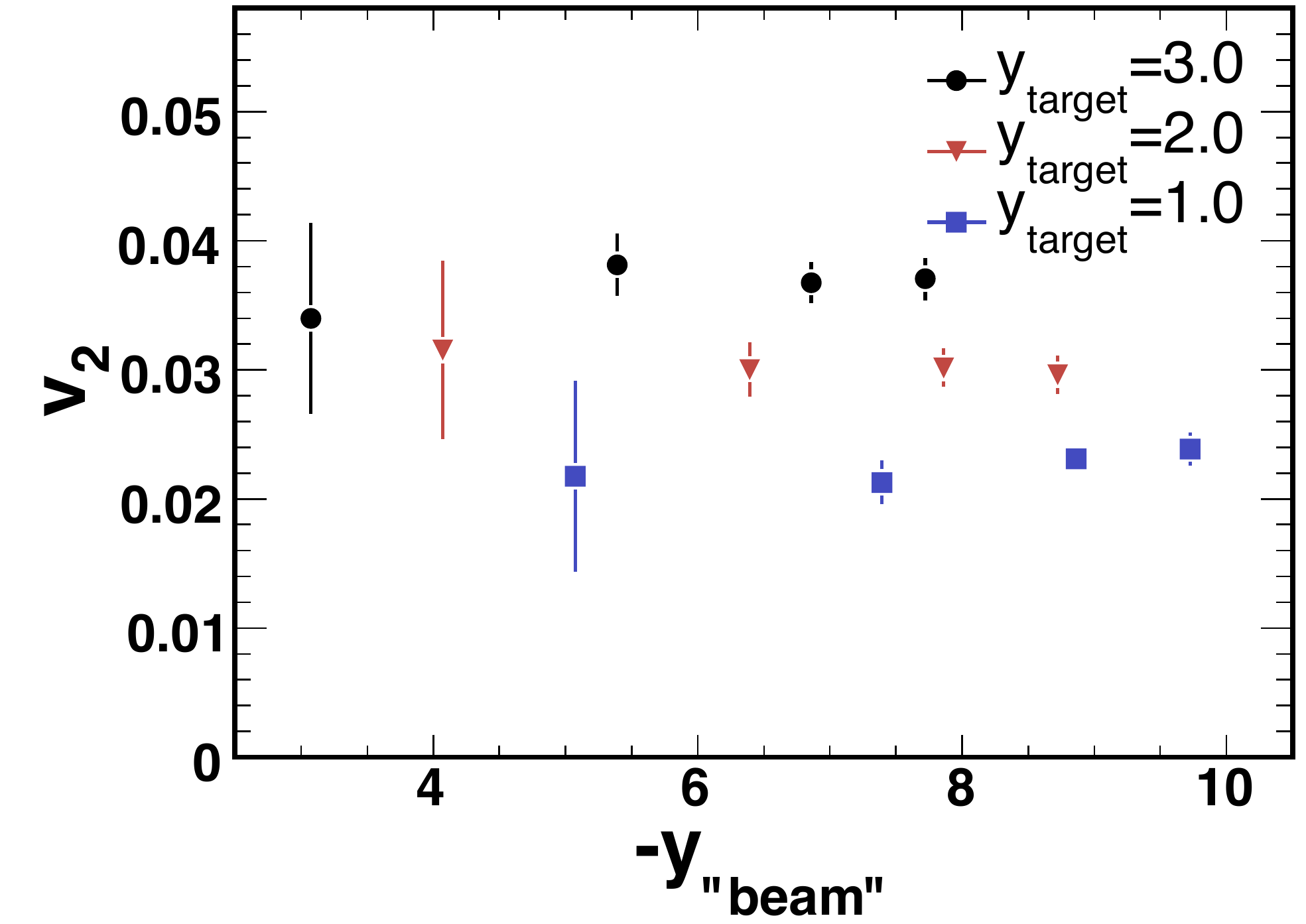} 
   \caption{(color on line)The elliptic flow parameter $v_2$ that would be measured in the ``gedanken experiment" (see text) if at $y=0$, in the laboratory frame of references, a Au beam with a rapidity $y_{``beam"}$ collided with a Au beam with rapidity $y_{``target"}$.  The data show that, for a given value as $y_{``target"}$, $v_2$ reaches a saturated value as $y_{``beam''}$ increases.  Furthermore that this saturated value is higher the larger is the value of $y_{``target"}$.  The figure is for \cite{11}.}
   \label{fig:9}
\end{figure}
\begin{figure}[htbp] 
   \centering
   \includegraphics[width=3.5in]{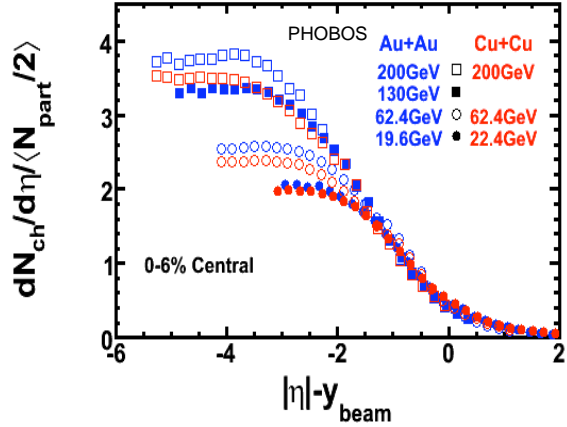} 
   \caption{(color on line) Compilation of PHOBOS data on the charged particle density $\frac{dN_{ch}}{d\eta}$ per participant pair produced in A+A collisions at various energies.  The data are plotted in the rest frame of one of the nuclei.  In boosting the data from the A+A center of mass system to he rest frame of one of the nuclei, the difference in pseudorapidity $\eta$ and rapidity $y$ is ignored.  The figure is from \cite{11}.}
   \label{fig:10}
\end{figure}

The results of the ``gedanken experiment" shown in figures 8 and 9 are the essence of this paper.  In words, they state that if, for a fixed value of $y_{``target"}$, we increase $y_{``beam"}$, from zero to very high values, at first the rapidity density of produced particles and the elliptic flow $v_2$, at $_{y=0}$, increase from zero until they reach a maximum value.  Further increases of $y_{``beam"}$, no matter how large, do not produce further increases of $\frac{dN}{d\eta}\mid_{ y=0}$ or of  $v_2\mid _{y=0}$.  In short, the ``potential'' of the ÒtargetÓ particle to produce more particles is saturated.  The only way to increase further $\frac{dN}{d\eta}\mid_{ y=0}$ and $v_2\mid _{y=0}$ is to raise the value of $y_{``target"}$.

Viewed in this light, we see that Extended Longitudinal Scaling is a direct manifestation of some kind of saturation taking place in high energy collisions.  It should be noted that the existence of a saturated ``potential" of the slower (``target") particle to produce particles which can only be increased by increasing its rapidity, is reminiscent of the Color Glass Condensate picture of high energy collisions \cite{12}.  As data at higher and higher energies from the LHC becomes available, it will be interesting to see if saturation continues to persist up to the highest energies.










\end{document}